\documentclass[iop,apjfonts]{emulateapj}
\newcommand{\DAMA}{{\sf DAMA}}
\newcommand{\CoGeNT}{{\sf CoGeNT}}
\newcommand{\CRESST}{{\sf CRESST}}
\newcommand{\CDMSSi}{{\sf CDMS-Si}}
\newcommand{\CDMSGe}{{\sf CDMS-Ge}}
\newcommand{\XENON}{{\sf XENON100}}
\newcommand{\LUX}{{\sf LUX}}

\usepackage{graphicx}
\usepackage{amssymb}
\usepackage{amsmath}

\usepackage{natbib}
\usepackage{hyperref}
\hypersetup{colorlinks=true,linkcolor=red,citecolor=blue,filecolor=cyan,urlcolor=magenta}
\usepackage{etoolbox}
\usepackage{multirow}
\usepackage{subfigure}

\begin{document}
 
\title[Helioseismology with long range dark matter-baryon interactions]{
Helioseismology with long range dark matter-baryon interactions}
\author{Il\'\i dio Lopes~\altaffilmark{1,2,7}, Paolo Panci~\altaffilmark{3,4,7}, 
Joseph Silk~\altaffilmark{4,5,6,7}}

\altaffiltext{1}{Centro Multidisciplinar de Astrof\'{\i}sica, Instituto Superior T\'ecnico, 
Universidade Tecnica de Lisboa , Av. Rovisco Pais, 1049-001 Lisboa, Portugal} 
\altaffiltext{2}{Departamento de F\'\i sica,Escola de Ciencia e Tecnologia, 
Universidade de \'Evora, Col\'egio Luis Ant\'onio Verney, 7002-554 \'Evora - Portugal} 
\altaffiltext{3}{CP3-Origins and DIAS, University of Southern Denmark, Odense, Denmark}
\altaffiltext{4}{Institut d'Astrophysique, UMR 7095 CNRS, Universit\'{e} Pierre et Marie Curie, 98bis Blvd Arago, 75014 Paris, France} 
\altaffiltext{5}{Department of Physics and Astronomy, The Johns Hopkins University, Baltimore, MD21218} 
\altaffiltext{6}{Beecroft Institute for Cosmology and Particle Astrophysics, University of Oxford, Keble Road,
Oxford OX1 3RH, UK} 
\altaffiltext{7}{E-mails: (IL) ilidio.lopes@tecnico.ulisboa.pt;
(PP) panci@iap.fr; (JS) silk@astro.ox.ac.uk} 

% % % % % % % % % % % % % % % % % % % % % % % % % % % % % % % % % % % % % % %
\date{\today}

\begin{abstract} 
Assuming the existence of a primordial asymmetry in the dark sector, we study how  DM-baryon long-range interactions, induced by the kinetic mixing of a new $U(1)$ gauge boson and the photon, affects the evolution of the Sun and in turn the sound speed profile obtained from helioseismology. Thanks to the explicit dependence on the exchanged momenta in the differential cross section (Rutherford-like scattering), we find that dark matter particles with a mass of $\sim 10\;{\rm GeV}$,  kinetic mixing parameter of the order of  $10^{-9}$ and a mediator with a mass smaller than a few MeV  improve  the agreement between the best solar model and the helioseismic data without being excluded by direct detection experiments. In particular, the \LUX\ detector will soon be able to either  constrain or confirm our best fit solar model in the presence of a  dark sector with long-range interactions that reconcile helioseismology with thermal neutrino results.
\end{abstract}

% insert suggested PACS numbers in braces on next line
%\pacs{ln}
% insert suggested keywords - APS authors don't need to do this

%\maketitle must follow title, authors, abstract, \pacs, and \keywords
\maketitle
% 
%%%%%%%%%%%%%%%%%%%%%%%%%%%%%%%%%%%%%%%%%%%%%%%%%%%%%%%%%%%%%%%%%%%%%%%%%%%%%
%
%%%%%%%%%%%%%%%%%%%%%%%%%%%%%%%%%%%%%%%%%%%%%%%%%%%%%%%%%%%%%%%%%%%%%%%%%%%%%
\section{Introduction}
%%%%%%%%%%%%%%%%%%%%%%%%%%%%%%%%%%%%%%%%%%%%%%%%%%%%%%%%%%%%%%%%%%%%%%%%%%%%%

%- dark matter large scale structure (observation and theory) 
%- dark matter small structure
The standard $\Lambda$CDM cosmological model has been 
successfully applied to explain the main characteristics of the Universe~\citep[see e.g.][]{2012arXiv1212.5226H, 2013arXiv1303.5076P}.  
In particular,  numerical simulations of collisionless cold Dark Matter (DM) 
describe the gravitational growth of  infinitesimal primordial density perturbations,
probed by the cosmic microwave background anisotropies,  
that lead to the formation of the present-day large-scale structure of the universe.  
These simulations  provide us with a number of predictions 
about the structure of cold DM halos and 
their basic properties~\citep[see e.g.][]{2013MNRAS.428.1351G}.
Although the standard cosmological model has proven
to be highly successful in explaining the observed large-scale  structure of the universe, it has
been less successful on  smaller scales. Recent data on low mass galaxies suggests 
that the inferred subhalo DM distributions have almost flat cores~\citep[see e.g.][]{2010AdAst2010E...5D}, 
in contradiction with the cuspy profile distributions predicted 
by numerical simulations~\citep[see e.g.][]{2010MNRAS.402...21N},  and that sub halo numbers are overpredicted at both low mass and intermediate masses~\citep[see e.g.][]{1999ApJ...522...82K} and massive dwarf galaxy scales~\citep[see e.g.][]{2013MNRAS.433.3539G}.

Some of the issues associated with dwarf galaxies can be addressed if the DM is more "collisional" (with baryons) 
than currently considered in numerical simulations.
{This hypothesis favors constant-density cores with lower central densities than those coming from cold DM 
models, such as self interacting DM~\citep[see e.g.,][]{2013MNRAS.430...81R}.}

In addition, the closeness between $\Omega_{\rm DM}$ and $\Omega_{\rm b}$, 
usually referred as cosmic coincidence, may suggest a  profound link between the dark  and the ordinary sectors. Indeed, although the two sectors have different macroscopic properties, the total amount of DM observed today could be  produced  in the early universe by a mechanism identical to baryogenesis, and therefore  an asymmetry between DM particles and their antiparticles would be expected. A detailed account of current progress in asymmetric DM studies can be found in the literature~\citep[see e.g.,][]{2013arXiv1305.4939P}.

\medskip
All of these cosmological facts may suggest that  dark and ordinary matter may have  more properties in common than expected. In view of this, it is    tantalizing to imagine that the dark world could be similarly complex (CP violating and asymmetric) and full of forces that are invisible to us (hidden parallel sector or mirror world\footnote{The idea of a mirror world was suggested prior to the advent of the Standard Model (see e.g. Refs.~\citep{Lee:1956qn,KobzarevOkunPomeranchuk}. The idea that the mirror particles might instead constitute the DM of the Universe was  discussed in refs.~\citep{Blinnikov:1982eh, 1983SvA....27..371B}.}).  The history of the early mirror universe has been studied in ref.~\citep{Berezhiani:2000gw}, while the impact of an hidden mirror sector in the CMB and LSS data can be found in ref.~\citep{Berezhiani:2003wj}. For a general review on the properties of a hidden world neighboring our own, see e.g.~refs.~\citep{Foot:2004pa, Berezhiani:2005ek}.

More specifically, since in our sector only long--range electromagnetic force and gravity affect the dynamical evolution of  virialized astrophysical objects, the physics of a complex dark sector in which the matter fields are charged under an extra $U(1)$ gauge group is particularly interesting to study. Indeed, if the mass of the new gauge boson (dark photon) is smaller that the typical  momenta exchanged in the scattering, the phenomenology of the dark world  can in itself be as complicated as that of our sector (e.g. dark electromagnetism with quite large self-interactions), providing at the same time also a feeble long-range interaction between the two worlds (thanks to the kinetic mixing of  the new $U(1)$ gauge boson with the photon). This implies in general that the dark world is more collisional than the standard cold DM, and therefore some of the previously mentioned problems can probably be solved~\citep[see e.g.][]{2011PhRvL.106q1302L}. 

In addition, since a long-range  DM-nucleus interaction is  enhanced for small  momentum exchanges, this class of models can also relax the tension between the positive results of direct detection experiments (the annual modulation observed by \DAMA~\citep{2008EPJC...56..333B,2010EPJC...67...39B} and \CoGeNT~\citep{2011PhRvL.107n1301A,2011PhRvL.106m1301A}, and the hints of an observed excess of events on \CRESST~\citep{Angloher:2011uu} and \CDMSSi~\citep{2013arXiv1304.4279C,Agnese:2013cvt}), and the constraints coming from null results (e.g.~\CDMSGe~\citep{2010Sci...327.1619C}, \XENON~\citep{2011PhRvL.107m1302A} and very recently \LUX~\citep{Akerib:2013tjd}). The phenomenology of long-range DM-nucleus interactions in direct DM searches has been studied in Refs.~\citep{2004PhRvD..69c6001F, Foot:2008nw, Foot:2010rj, Foot:2012rk,2011PhRvD..84k5002F, Panci:2012qf}.

\medskip
In this paper,  we investigate how a DM-baryon Long Range Interaction (DMLRI), induced by the kinetic mixing 
of a new $U(1)$ gauge boson and the photon, affects the evolution of the Sun. The results obtained 
are then confronted with helioseismology data. The helioseismic data that we use was obtained by 
several international collaborations,  such as the  Solar and Heliospheric Observatory (SOHO) mission 
and the  Birmingham Solar Oscillations Network (BiSON) observational
network~\citep{1997SoPh..175..247T,2009ApJ...699.1403B}.
Furthermore, we also discuss the constraints on the main parameters 
of long-range DM particle interactions which can be obtained from direct 
DM search data. 
%%%%%%%%%%%%%%%%%%%%%%%%%%%%%%%%%%%%%%%%%%%%%%%%%%%%%%%%%%%%%%

%%%%%%%%%%%%%%%%%%%%%%%%%%%%%%%%%%%%%%%%%%%%%%%%%%%%%%%%%%%%%%%%%%%%%%%%%%%%%
\section{Properties of DM with long-range interactions}\label{Sec2}

\begin{figure}
\centering
\includegraphics[scale=.9]{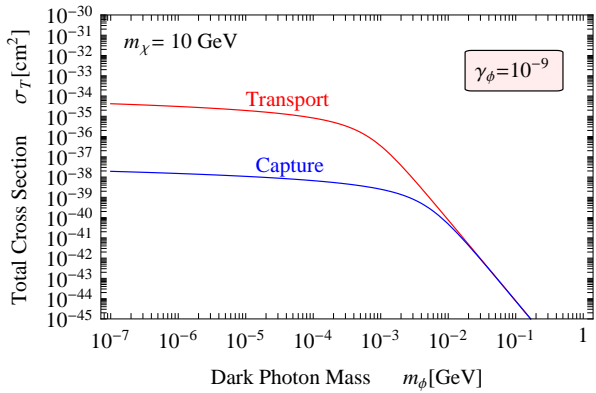}
\caption{An illustrative example of the DM-hydrogen energy transfer cross section 
as a function of the mass of the dark photon $m_\phi$ for a fix value of the kinetic mixing parameter $\gamma_\phi=10^{-9}$ and a DM mass $m_\chi=10$ GeV. In blue is shown the scattering cross section responsible for the capture process $\sigma_T^{\rm cap}$, while in red the one entering in the energy transport mechanism $\sigma_T^{\rm tra}$ computed at the present time ($T_c^0=1.57\times 10^7$ K).}
\label{fig:1}
\end{figure}

%%%%%%%%%%%%%%%%%%%%%%%%%%%%%%%%%%%%%%%%%%%%%%%%%%%%%%%%%%%%%%%%%%%%%%%%%%%%%
%      
% - General motivation for the Mirror DM and discuss this particular case....
% - how different his this model from others....
%
In most of the classical models, one often assumes a symmetric dark  sector in
which the DM scattering off of baryon nuclei is done by a contact-like interaction. 
In our study, we will instead focus on a class of asymmetric DM models in which the interaction 
between DM particles and target nuclei is mediated by a light messenger. If the typical 
 momenta exchanged in the scattering is bigger than the mass of the mediator, a long--range
interaction then arises.

\medskip
A specific realization of this kind of picture is offered by particle physics models where a new $U(1)$ hidden gauge boson $\phi$  (dark photon)
possesses a small kinetic mixing $\epsilon_\phi$ with the photon. 
In this case, the interaction between a nucleus with mass $m_T$ and electric charge $Ze$ 
($Z$ is the number of protons in the baryon nucleus  and $e$ the electrical charge) and 
a DM particle with mass $m_\chi$ and dark charge $Z_\chi g_\chi$ ($Z_\chi $ and $g_{\chi}$ are the equivalent quantities of $Z$ and $e$ in the dark sector) is described in the non-relativistic limit by the following 
%Yukawa potential~\citep[see e.g.][]{2010PhRvL.104o1301F, 2011PhRvD..84k5002F},
{Yukawa potential~\citep[see e.g.,][]{2011PhRvD..84k5002F}},  
\begin{equation}
\label{Vr}
V(r) = \epsilon_\phi \, k_\chi \, \frac{Z \alpha}r \,e^{-m_\phi r} \ ,
\end{equation}
where  $\alpha = e^2/4\pi$ is the electromagnetic fine structure constant 
and the parameter $k_\chi=Z_\chi g_\chi/e$ measures the strength of the DM-dark photon coupling. Here $m_\phi$ is the mass of the dark photon that acts like an electronic cloud which screens the charges of the particles involved in the scattering. Since both $k_\chi$ and $\epsilon_\phi$ are unknown, we define $\gamma_\phi\equiv k_\chi \, \epsilon_\phi$ and we treat it as a free parameter of our model together with $m_\chi$ and $m_\phi$. {Before moving on we have however verify whether the symmetric component of $\chi$ in the dark sector is annihilated away or not. In general, since for this simple model the dominant annihilation channel is given by $\chi \chi \rightarrow \phi \phi$, we can put a lower bound on the parameter $k_\chi$ by demanding that the annihilation cross section into dark photons ($\langle \sigma v \rangle_{\phi\phi}= \pi \alpha^2 k_\chi^4 /(2 \, m_\chi^2) $) is bigger than the thermal one ($\langle \sigma v \rangle_{\phi\phi}\sim 1$ pb). In agreement with Refs.~\citep{Tulin:2013teo,Kaplinghat:2013kqa}, this condition requires $k_\chi \geq k_\chi^\Omega = \bar k_\chi^\Omega \sqrt{m_\chi/{\rm GeV}}$, where $\bar k_\chi^\Omega  \simeq 7.5 \times 10^{-2}$. We shall come back to this point at the end of Sec.~\ref{Sec:3} because for this kind of models the DM-DM scattering can  plays an important role in the capture rate and in the energy transport by DM particles in the Sun.}

\medskip
The differential cross section, neglecting the form factor of the target nuclei\footnote{The Sun being mostly composed of hydrogen and helium, we can  justify neglecting the nuclear responses. On the other hand for the derivation of the direct detection constraints, we use the form factors provided in ref.~\citep{Fitzpatrick:2012ix}.}, can be simply obtained by performing the Fourier transform of Eq.~\ref{Vr} and it reads
\begin{equation}\label{diffCC}
\frac{{\rm d}\sigma_T}{{\rm d}\Omega}= \frac{\xi_\chi^2\mu^2}{(q^2+m_\phi^2)^2}\ ,
\end{equation}
where $\xi_\chi = 2 \alpha Z \, \gamma_\phi$, $\mu=m_\chi m_T/(m_\chi + m_T)$ is the DM-nucleus reduced mass and $q=\sqrt{2 m_T E_{\rm R}}$ is the  momenta exchanged in the interaction with $E_{\rm R}$ the recoil energy. As is apparent from the dependence on the dark photon mass,  two different regimes clearly appear:
\begin{itemize}
\item[$\diamond$] 
\emph{Point--like limit} ($q\ll m_\phi$): In this regime the interaction is of a contact type. Indeed the differential cross section turns out to be proportional to $\xi_\chi^2  /m_\phi^4$ which plays the same role as  Fermi's constant in weak interactions. The interaction reduces to the ``standard'' spin-independent picture, apart from the fact that in this case, DM particles only couple with protons (${\rm d}\sigma_T/{\rm d}\Omega \propto Z^2$).

\item[$\diamond$] 
\emph{Long--range limit} ($q\gg m_\phi$): In this regime,  the differential cross section acquires an explicit dependence on the momenta exchanged in the interaction and therefore a Rutherford-like cross section arises (${\rm d}\sigma_T/{\rm d}\Omega \propto 1/q^4$). This is extremely interesting because ideal experiments with very low energy threshold and light target nuclei (e.g.~the Sun) are in principle more sensitive than the ones with high threshold and heavy targets (e.g.~\XENON\ and \LUX). To give a concrete example, once  DM particles are thermalized with baryons in the center of the Sun, their collisions occur with $q\simeq 1$ MeV considering a DM mass of 10 GeV. For direct detection experiments, the typical momenta transferred in the scattering is instead bigger ($q\gtrsim 20$ MeV). Thanks to this fact, we expect that DM models feature a long-range interaction with ordinary matter can affect the sound-speed radial profile  of the Sun 
without being excluded by terrestrial experiments. 

\end{itemize}
 
{Having  the differential cross section at our disposal, we assume that the normalization of both the capture rate and the  transport of energy by DM particles in the Sun is controlled by the energy transfer cross section obtained by weighting Eq.~\eqref{diffCC} with ($1-\cos\theta$). This is of course a good estimator for the transport mechanism, while for the capture process the introduction of the same cross section  is justified because the Sun's escape velocity $v_{\rm esc}(r)$ is much bigger than both the thermal velocity of baryons in the Sun and the typical dispersion velocity of the DM particles in the halo $v_0$. In particular, we have checked that the error in the total rate is negligible,  if we replace the total cross section  in the rate per unit time $\Omega(w)$ given e.g. in Eq.~7 of~\citep{Kumar:2012uh},  with the energy transfer cross section computed for $w=v_{\rm esc}(r)$.  In the Born approximation\footnote{The Born approximation is valid if $\beta_\phi \lesssim 0.1$ (see e.g.~\cite{2011PhRvL.106q1302L}). Since we are interested in the long-range regime ($\mu v_{\rm rel}^2 \gg m_\phi$), it is easy to check that this approximation is very well satisfied.}, the energy transfer cross section writes,
\begin{eqnarray}
\sigma_{T}(v_{\rm rel})= \frac{2\pi \beta_\phi^2}{m_\phi^2}
\left[\ln{\left(1+r_\phi^2\right)}-\frac{r_\phi^2}{1+r_\phi^2}  \right] \ ,
\label{eq-sigma}
\end{eqnarray}}
where  $\beta_\phi=\xi_\chi m_\phi/ (2\mu  v^2_{\rm rel}) $,  $r_\phi= 2\mu v_{\rm rel}/m_\phi$,  and
$v_{\rm rel}$ is the relative velocity between the DM flux and the Sun. Unlike the customary DM 
models, in this case  $\sigma_{T}$ depends on $v_{\rm rel}$ in the long--range regime. Thanks to this main novelty,  one therefore
expects that the typical scattering cross section in the capture process differs compared to the one 
entering in the transport mechanism. On a more specific level, one has:

\begin{itemize}
{\item[$\diamond$] \emph{Capture:} In general, the infalling DM particles reach a given 
shell of radius $r$ with a velocity $w(r)=\sqrt{u^2+v_{\rm esc}^2(r)}$, where $u$ is the DM velocity at infinity 
with respect to the Sun's rest frame. Since both the thermal
velocity of baryons in the Sun and the dispersion velocity of DM particles in the halo  are 
much smaller than $v_{\rm esc}$, we can assume as commented upon above 
 that the relative velocity $v_{\rm rel}\equiv w(r) \simeq v_{\rm esc}(r)$. }
Furthermore, since  the total number of DM particles captured by the Sun is independent on $r$, it is a good 
approximation to define an  average infalling DM velocity by 
{\begin{equation}
\bar w=\frac1{M_\odot} \int_0^{R_\odot} {\rm d}^3 r \, w(r) \, \rho(r) \simeq 1120 \mbox{ km/s}\ .
\end{equation} 
Here $\rho(r)$ is the Sun's mass density, $M_\odot=\int {\rm d}^3 r \, \rho(r)\simeq 1.98\times 10^{30}$ kg is its total mass and $R_\odot\simeq 6.95 \times 10^{10}$ cm is its radius.}
% as most of the Sun's mass is concentrated within 0.1 solar radius  it is a good approximation to define 
%an averaged infalling velocity given by $\bar w\equiv w(0.1\,r_\odot)\simeq 1315$ km/s. 
%For instance, considering a typical velocity dispersion $v_0=220$ km/s, we get $\bar w\simeq 1120$ km/s. 
As explained in more detail  in the next section,  the capture rate is then computed numerically considering a 
constant cross section $\sigma_T^{\rm cap}=\sigma_T(\bar w)$.

\item[$\diamond$] \emph{Transport:} In this case, the typical relative velocity for the scattering is much smaller than $\bar w$ being due to DM particles thermalized together with the ordinary plasma in the center of the Sun.  It is then a good approximation  to assume $v_{\rm rel}\equiv v_{\rm th}$, where $v_{\rm th}=\sqrt{2 T_c/m_\chi}$ is the thermal speed and $T_c$ is the time-dependent temperature in the Sun's core. As explained in more detail in the next section, we compute  the transport of energy numerically by considering $\sigma_T^{\rm tra}=\sigma_T(v_{\rm th})$. It is worth stressing that since the solar code follows the time evolution  of the Sun, in the early stages  the energy transport, and in turn the thermal conduction by DM particles,  was much more efficient with  $T_c$ at that time being smaller than the present-day central temperature. 
 \end{itemize}

In this study, we focus on the interaction of DM with hydrogen
-- the most abundant chemical element in the Sun's interior.
Fig.~\ref{fig:1} shows an illustrative example of the DM-hydrogen energy transfer cross section $\sigma_{T}$ as a function of the mass of the
dark photon $m_\phi$ for a fixed value of the kinetic mixing parameter $\gamma_\phi=10^{-9}$ and a DM mass $m_\chi=10$ GeV. On a more specific level we   show in blue the scattering cross section responsible for the capture process $\sigma_T^{\rm cap}$, while in red is shown the one entering into the energy transport mechanism $\sigma_T^{\rm tra}$ computed at the present time ($T_c^0=1.57\times 10^7$ K). We can see that if the mass of the dark photon is smaller than a few MeV (long--range regime), the capture and the transport processes are controlled by different scattering cross sections. It is worth pointing out that in this limit,  the ratio $\sigma_T^{\rm tra}/\sigma_T^{\rm cap}$ is barely dependent  on $m_\phi$, if $m_\chi$ is larger than the hydrogen mass. It instead depends on the mass of the DM particle through the thermal velocity relation and in particular for a 10 GeV candidate, $\sigma_T^{\rm tra}\sim 10^3\,\sigma_T^{\rm cap}$. The main new aspect is actually given by this enhanced conduction in the inner part of the Sun compared to the usual DM models. Thanks indeed to this fact, DM particles, interacting via long-range forces with ordinary matter, can produce an impact on the helioseisomology data without evading the constraints coming from direct DM search experiments. 

\medskip
In our analysis, we consider the observed sound speed radial profile of the Sun and compare it to the theoretical prediction over a broad range of DM masses (4 GeV$\leq m_\chi\leq$ 20 GeV), {dark photon masses (0.1 keV $\leq m_\phi\leq$ 1 GeV)\footnote{As we will see later, for masses below 0.1 keV the Sun's sound speed profiles become independent on $m_\phi$ and therefore our results can also  be extended for lighter mediators.} and kinetic mixing parameters ($10^{-12}\leq \gamma_\phi \leq10^{-6}$). With these choices, the scattering  cross-section spans a large interval  of values from 
$5\times 10^{-29}\;{\rm cm^2 }$ to  $8\times 10^{-55}\;{\rm cm^2 }$. 
The cross-section range of interest for the Sun corresponds to 
the values for which $ \sigma_{T} $ is  close to  the Sun's characteristic scattering  cross-section, 
$\sigma_{\odot}= m_p/M_{\odot } \;R_{\odot}^2$. 
From the zero-age mean-sequence (ZAMS) until the present age of the Sun ($4.6\;{\rm Gyear}$),  
$\sigma_{\odot}$ takes values between  $2.5\times 10^{-32}$ cm$^2$ and $4\times 10^{-36}$ cm$^2$.}

\begin{figure}
\centering
\includegraphics[scale=.9]{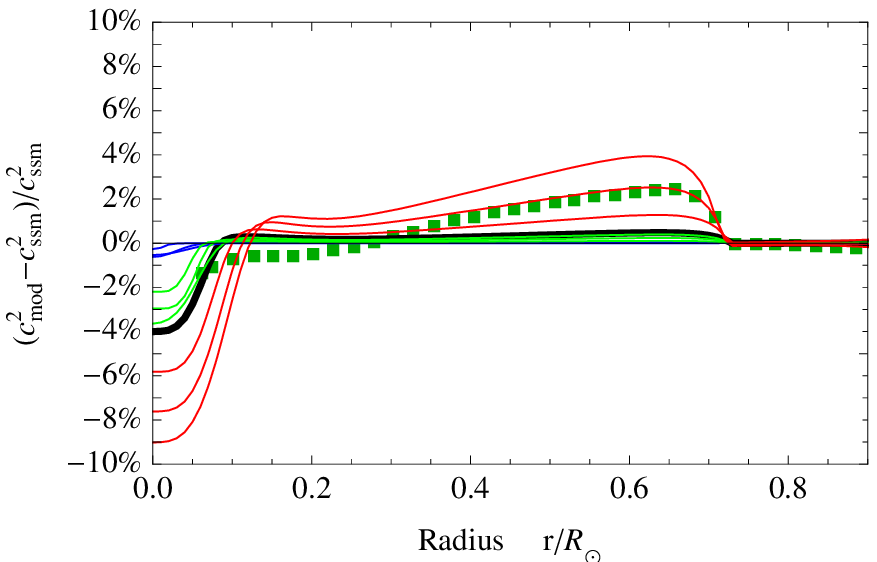} 
\caption{
{Comparison of sound speed differences:  
$\Delta c^2 =(c_{\rm mod}^2 - c^2_{\rm ssm})/c^2_{\rm ssm} $, 
$c^2_{\rm ssm}$ is the sound speed of the SSM~\citep[e.g.,][]{2013ApJ...765...14L}
and $c^2$ is either $c^2_{\rm obs}$ the observed sound speed~\citep[green-square dots:][]{1997SoPh..175..247T,2009ApJ...699.1403B} or $c^2_{\rm mod}$
the sound speed of a DMLRI solar model 
(continuous curves: [$|\Delta c^2| < 2.0 \%  $] blue, 
[$2.0 \le |\Delta c^2| \le 4.0 \%$]  green  and [$|\Delta c^2| > 4.0 \%$] red).
The DMLRI models were computed for the parameters (see text): 
$5 \; {\rm GeV} \leq m_\chi \leq20 \; {\rm GeV} $;  $0.1\; {\rm keV} \leq  m_\phi \leq 1 \; {\rm GeV} $ and $\gamma_\phi=10^{-9}$.  
The black curve corresponds to a fiducial model with $m_\chi=10\;{\rm GeV}$ and $m_\phi=10\;{\rm keV}$. 
Note that the observational error in $c_{\rm obs}$ is multiplied by a factor 10. 
}}
\label{fig:2}
\end{figure}

%%%%%%%%%%%%%%%%%%%%%%%%%%%%%%%%%%%%%%%%%%%%%%%%%%%%%%%%%%%%%%%%%%%%%%%%%%%%%
\section{Dark matter and the Sun}
\label{Sec:3}
%%%%%%%%%%%%%%%%%%%%%%%%%%%%%%%%%%%%%%%%%%%%%%%%%%%%%%%%%%%%%%%%%%%%%%%%%%%%%

The Sun, as are all stars in the Milky Way, is immersed in a halo of DM.
As with any other star in the galaxy, the Sun captures substantial numbers 
of DM particles during its evolution, but     
any impact on the star depends on the properties of these particles,  
as well as on the stellar dynamics and structure of the star.
In general, for low-mass stars evolving in low density DM halos,
the presence of DM inside the star changes its evolution 
by providing the star  with a new mechanism to evacuate the heat
produced in the stellar core~\citep[e.g.,][]{2002MNRAS.331..361L,2002PhRvL..88o1303L,2011PhRvD..84j1302Z,2012ApJ...757..130L,2013ApJ...765L..21C}.
This is quite different for stars evolving in DM halos of high density -- 
as occurs during the formation of the first generation of stars. 
In these cases, the annihilation of DM particles supplies the star with
an additional source of energy capable of substantially extending the
lifetimes of these stars~\citep{2009MNRAS.394...82S,2009ApJ...705..135C,2011PhRvD..83f3521L,2011ApJ...742..129S}.

The computation of the impact of DM in the evolution of the Sun is done
by a modified version of the {\sc cesam} code~\citep{1997A&AS..124..597M,2008Ap&SS.316...61M}, 
which has been widely used to compute the SSM and for modelling other stars 
by different research groups~\citep{2010A&A...514A..31D,2010ApJ...715.1539T,2013arXiv1308.3346L,2013ApJ...765...14L}.
In this study, we follow a procedure identical to other studies published 
in the literature by some of us as well as other authors~\citep[e.g.,][]{2002MNRAS.331..361L, 2002PhRvL..88o1303L, 2010PhRvD..82j3503C, 2010PhRvD..82h3509T, 2010Sci...330..462L, 2010ApJ...722L..95L, 2012ASPC..462..537H, 2012ApJ...757..130L, 2012ApJ...752..129L,2014ApJ...780L..15L}.
Nevertheless, there are some important differences between this study and the  
previous ones, which we will highlight in the remainder of this section. 
 
\medskip
As commented upon in the previous section, the impact of DM in the 
Sun's interior depends on two major physical processes: 
the accumulation of DM inside the star, and the efficiency of DM 
in  transferring energy from the core to the external layers. 
In any case, the density of DM in the halo is the single major 
ingredient affecting  the impact of DM in the Sun. 
As usually done in these studies, 
we  consider that the local density of DM   is  
$\rho_\odot=0.38 \; {\rm GeV/cm^3} $~\citep{1995ApJ...449L.123G,1996PhRvD..53.4138G,2010JCAP...08..004C,Salucci:2010qr}.
The choice of this value is in part made to facilitate the comparison with other work.  
Still, this is a very reliable value: the most recent estimates of $\rho_\odot$ made by two  
independent groups have obtained values of  $ 0.3 \; {\rm GeV / cm^3} $~\citep{2012ApJ...756...89B}
and $ 0.85 \; {\rm GeV/cm^3} $~\citep{2012MNRAS.tmp.3493G}. In particular~\citep{2012MNRAS.tmp.3493G} 
argue that their new method is quite robust, and they have obtained their value at 90\% confidence level.
In our computation, we will also consider that the DM particles in the solar neighborhood 
have dispersion velocity  $v_0=220 \; {\rm km / s}$~\citep[see e.g.,][]{2005PhR...405..279B}.
   
\medskip    
The accumulation of DM in the Sun's core during its evolution
from the beginning of the ZAMS until the present age ($4.6\; {\rm Gyear}$) 
is regulated by three physical processes: capture, 
annihilation and evaporation of DM particles. 
At each step of the evolution, the total number of DM particles $N_\chi$,
that is captured by the star  is given by  
{\begin{equation}
\frac{dN_\chi (t)}{dt}=\Gamma_{\rm cap}+\Gamma_\chi N_\chi(t)-\Gamma_{\rm ann} N_\chi(t)^2 -\Gamma_{\rm eva}N_\chi(t),
\label{eq-N}
\end{equation}
where $\Gamma_{\rm cap}$, $\Gamma_\chi $, $\Gamma_{\rm ann}$ and $\Gamma_{\rm eva}$
refer  to the DM capture, self-capture, annihilation and evaporation rates respectively. }
It is worth noticing that, unlike  previous studies, 
in this work we resolve numerically the equation (\ref{eq-N}) for
each step of the star's evolution.  A detailed discussion about these processes 
can be found in the literature~\citep[e.g.,][]{1987NuPhB.283..681G,2011PhRvD..83f3521L}.

The capture rate $\Gamma_{\rm cap}$  is computed numerically  
from the expression obtained by~\citet{1987ApJ...321..571G} 
as implemented by~\citet{2004JCAP...07..008G}. The scattering of
DM particles with the baryons inside the Sun is the main
factor affecting the capture rate $\Gamma_{\rm cap}$. We restrict our study 
to the scattering  of DM particles to the most abundant element, i.e., hydrogen.
We consider that the capture rate is 
controlled by $\sigma_T^{\rm cap}$ obtained by substituting $v_{\rm rel}$ with the averaged infalling speed $\bar w$ in Eq.~(\ref{eq-sigma}) as briefly explained in the previous section. In particular, in the long-range regime
($r_\phi\gg 1$ or in terms of the exchanged momenta, $q\gg m_\phi$), the DM-hydrogen energy transfer cross section responsible for the capture process is independent of $m_\chi$ and logarithmically  dependent on $m_\phi$. It reads
\begin{equation}\label{eq-sigma-coll-LR}
\begin{split}
& \lim_{r_\phi\gg1} \sigma_T^{\rm cap} =  4\pi \, \frac{ \alpha^2 \gamma_\phi^2}{ m_p^2 \bar w^4} L_\phi^{\rm cap}(m_\phi) \simeq \\
&  \left(\frac{\gamma_\phi}{10^{-9}}\right)^2\left(\frac{L^{\rm cap}_\phi}{\bar L_\phi^{\rm cap}}\right) \cdot 1.1\times 10^{-38}\mbox{ cm}^2\ ,
\end{split}
\end{equation}
where $L_\phi^{\rm cap}=\ln(2 m_p \bar w/m_\phi)$ is a sort of Coulomb logarithm that measures the strength of the screening effect in the capture process and $\bar L_\phi^{\rm cap}\simeq 6.5$ is its value  for $m_\phi=10$ keV. 
 
The other chemical elements are  under-abundant,  consequently 
their contribution for the capture of DM is negligible.  
The dependence of the scattering cross-section on the parameter space of
long-range DM particles is given by equation~(\ref{eq-sigma}).
The description of how this capture process is implemented 
in our code is discussed in~\citet{2011PhRvD..83f3521L}.

\medskip
{Since we assume a primordial asymmetry between particles and anti-particles in the dark sector, the annihilation rate $\Gamma_{\rm ann} $ is set to be zero. This condition, as we have briefly mentioned in Sec.~\ref{Sec2} is justified if  $k_\chi \geq k_\chi^\Omega \simeq \bar k_\chi^\Omega \sqrt{m_\chi/{\rm GeV}}$.}

\medskip
The evaporation rate $\Gamma_{\rm eva}$ is relevant only for very light particles, 
i.e., particles with $m_\chi \le 4\; {\rm GeV}$~\citep{1990ApJ...356..302G}.
\citet{2011NuPhB.850..505K} estimated the evaporation 
mass for DM particles in the Sun,  $m_{\rm eva}$, to be 
such that $m_{\rm eva}=3.02+0.32\log_{10}{(\sigma_T/10^{-40} {\rm cm^2})}\;{\rm GeV}$.
If $m_\chi \le m_{\rm eva} $ the DM particle escapes
the solar gravitational field and consequently has no impact on the structure of the star. 
\citet{2011NuPhB.850..505K} also  found that the evaporation of DM particles is
completely irrelevant for $m\ge 8\;{\rm GeV}$.
It is worth noticing the fact that because the evaporation boundary 
has a exponential dependency on the  mass and the scattering cross-section 
of the DM particle~\citep{1990ApJ...356..302G}, 
it follows that if $m_\chi$ exceeds $m_{\rm eva}$ by a few percent,
then the evaporation of DM particles is totally negligible.
Furthermore it is also important to point out that  our DM model can
easily have a quite large self--interaction in the long--range regime with size similar
to the electromagnetic scattering. If then this self--interaction is
attractive (e.g. this can be obtained if the dark sector is composed by light and heavy 
species with a different sign of the dark charge $Z_\chi g_\chi$) we expect that $\Gamma_{\rm eva}$ can be set to zero  for DM masses below 4 GeV as well, in such cases the properties of the dark plasma being similar to the ordinary one 
(electrons are indeed trapped in the Sun). 
We leave further discussion of  this  important effect for future studies.

In our computation, we use  $\Gamma_{\rm eva}$ estimated by~\citet{2013JCAP...07..010B}
in the regime where the Sun is optically thin with respect to the DM particles.
Nevertheless, we restrict our analysis to particles with mass larger than $4\;{\rm GeV}$,  
for which the evaporation rate is almost negligible.

\medskip
Once gravitationally captured by the star, the DM particles 
thermalise with baryons after a few Kepler orbits around the solar centre,  
colliding through elastic scattering  with hydrogen and other elements,
and thus providing the star with an alternative mechanism for the transport of energy.  
The relatively efficiency of the DM energy transport in relation 
to the radiative heat transport depends on the Knudsen number, 
{$K_\chi=l_\chi/r_\chi$ where $l_\chi $ is the free mean 
path of the DM particle inside the star and $r_\chi (m_\chi)$ is the  characteristic 
radius of the DM distribution~\citep{1986ApJ...306..703G,2002MNRAS.331..361L} given by
\begin{equation}\label{eq-meanfreepath-LR}
\begin{split}
& r_\chi(m_\chi)=\left(\frac{3 T_c}{2\pi G_{\rm N} \rho_c\,m_\chi}\right)^{\frac12} \simeq \\
&  \left(\frac{10 \mbox{ GeV}}{m_\chi}\right)^{\frac12} \hspace{-.08cm}\left(\frac{T_c}{T_c^0}\right)^{\frac12}\hspace{-.08cm} \left(\frac{\rho_c^0}{\rho_c}\right)^{\frac12} \cdot 0.035 \, R_\odot \ ,
\end{split}
\end{equation}
where $\rho^0_c\simeq 8.3 \times 10^{25}$ GeV/cm$^3$ is the today's density of the Sun's core.}     
Depending on the value of $\sigma_T $, the transport of energy by DM 
is local (conductive) or non-local, which corresponds to  $K_\chi \ll 1$ or $K_\chi \gg 1$. {In the conductive regime the effective luminosity carried by DM particles is proportional to $n_\chi l_\chi$, while in the non-local one to $n_\chi / l_\chi$. The maximal luminosity is instead achieved when $l_\chi \simeq 10 \, r_\chi$ (see e.g. Fig.~11 of \cite{1990ApJ...352..654G}). Here $n_\chi$ is the number density of DM particles captured in the Sun which is given by $n_\chi(r)=  n_0\,\mbox{exp}[-r^2/r_\chi^2]$ where $n_0 = N_\chi/(\pi^{3/2}r_\chi^3)$ (see e.g.~\cite{2002MNRAS.331..361L}).} We assume that the thermal conduction by DM particles is controlled by $\sigma_T^{\rm tra}$ obtained by substituting $v_{\rm rel}$ with the thermal speed $v_{\rm th}$ in Eq.~(\ref{eq-sigma}) as briefly explained in the previous section. Unlike the cross section responsible for the capture
process, in the long--range limit $\sigma_T^{\rm tra}$ depends on the DM mass via the thermal velocity $v_{\rm th}$. It explicitly reads
\begin{equation}\label{eq-sigma-tra-LR}
\begin{split}
& \lim_{r_\phi\gg1} \sigma_T^{\rm tra} =  \pi \, \frac{ \alpha^2 \gamma_\phi^2\,m_\chi^2}{ m_p^2 T_c^2}L_\phi^{\rm tra}(m_\chi,m_\phi,T_c) \simeq \\
&  \left(\frac{m_\chi}{10\mbox{ GeV}}\right)^2 \hspace{-.08cm}\left(\frac{T_c}{T_c^0}\right)^2\hspace{-.08cm} \hspace{-.08cm} \left(\frac{\gamma_\phi}{10^{-9}}\right)^2\hspace{-.08cm} \left(\frac{L_\phi^{\rm tra}}{\bar L_\phi^{\rm tra}}\right) \cdot 1.9\times 10^{-35}\mbox{ cm}^2\ ,
\end{split}
\end{equation}
where $L_\phi^{\rm tra}(m_\chi,m_\phi,T_c)=1/2\ln(8 m_p^2 T_c/(m_\chi m_\phi^2))$ emulates the screening effect in the energy transport and $\bar L_\phi^{\rm tra}\simeq 4.5$ is its value for $m_\phi=10$ keV, $m_\chi=10$ GeV and $T_c=T_c^0$. {Once computed the energy transfer cross section the average mean free path in the long-range regime, can be estimated as
\begin{equation}\label{eq-meanfreepath-LR}
\begin{split}
& l_\chi = \left(\langle n_b \rangle \cdot \lim_{r_\phi\gg1} \sigma_T^{\rm tra}\right)^{-1}    \simeq \\
&  \left(\frac{10 \mbox{ GeV}}{m_\chi}\right)^2 \hspace{-.08cm}\left(\frac{T_c^0}{T_c}\right)^2\hspace{-.08cm} \hspace{-.08cm} \left(\frac{10^{-9}}{\gamma_\phi}\right)^2\hspace{-.08cm} \left(\frac{\bar L_\phi^{\rm tra}}{ L_\phi^{\rm tra}}\right) \cdot 0.055 \, R_\odot \ ,
\end{split}
\end{equation}
where 
\begin{equation}
\langle n_b \rangle=\frac1{R_\odot} \int_0^{R_\odot}\hspace{-.15cm} {\rm d} r  \, n_b(r) \simeq 1.3 \times 10^{25} / \mbox{cm}^3  \ ,
\end{equation}
is the average number density of baryons in the Sun and $n_b(r) \simeq \rho(r) / m_p $.} Depending on the free parameters of the model scanned in our analysis, the energy transfer cross section 
covers a  broad range of values ($5\times 10^{-29}\;{\rm cm^2}\leq \sigma_T^{\rm tra}\leq 8\times 10^{-55}\;{\rm cm^2}$). In this case the corresponding range of the Knudsen number  is ($ 5 \times 10^{-9}\leq K_\chi \leq 3 \times 10^{18}$). Both mechanisms of energy transport are considered in this study~\citep{1986ApJ...306..703G,2002MNRAS.331..361L}. In the case of the non-local regime,  we follow the numerical  prescription of~\citet{1990ApJ...352..654G} rather than the original one proposed by~\citet{1985ApJ...294..663S}. 

{We note that in the long-range limit, a Rutherford-like DM-baryon interaction  can significantly enhance the energy transport ($\sigma_T^{\rm tra} \gg \sigma_T^{\rm cap}$) compared to the usual picture (for the standard spin-independent and spin-dependent cases, one always has $\sigma_T^{\rm tra} = \sigma_T^{\rm cap}$). Indeed, even if the capture cross section is around $10^{-39}$ cm$^2$ (this choice correspond to $\gamma_\phi=3.3\times 10^{-10}$), the DM particles can have a mean free path around $10\,r_\chi$, providing then the maximum luminosity carried by DM particles. For the standard contact interaction and for the same cross section, the mean free path is much longer and therefore the luminosity is reduced being in this case the transport non-local.} This is the main new element, and in particular, as we will see in  Sec.~\ref{CConstraints},  this class of models with enhanced energy transport can solve the so-called solar abundance problem  without being excluded by direct searches experiments.

\medskip
A similar conclusion can also be found in Ref.~\citep{Vincent:2013lua}. In particular they implement the formalism of~\citet{1990ApJ...352..654G} in order to properly account for both velocity and exchanged momenta dependencies in the differential cross section. Although their method is more refined compare to the one we are using they  did not  incorporate their results in a solar simulation software yet.  

\medskip
{The self-capture rate $\Gamma_\chi$ (as in Eq.~\ref{eq-N}) is neglected in our computation. Nonetheless, since in the long-range regime the self-interaction between the DM particles  can be relatively large, we could expect a major impact in the Sun's evolution. As it will become clear later on we find that for  $k_\chi \sim k_\chi^\Omega$  the contribution coming from $\Gamma_\chi$ is negligible because the two leading processes (capture and transport) compensate each other. On the other hand for larger value of the parameter $k_\chi$ we expect that the self-interaction entirely dominates the transport of energy in the Sun and therefore the results presented in this paper will be no longer valid. In the following, we estimate the impact of including the self-capture in Equation~\eqref{eq-N}.

The starting point is the definition of the self-capture energy transfer cross section $\sigma_\chi^{\rm cap}$ that in the Born approximation can be computed  by substituting in Eq.~\ref{eq-sigma}, $\mu  \rightarrow m_\chi/2$ and $\gamma_\phi \rightarrow k_\chi$. Since the DM particles occupy a very small range of radii within the Sun, it reads 
\begin{equation}\label{eq-sigmaSI-coll-LR}
\begin{split}
& \lim_{r_\phi\gg1} \sigma_\chi^{\rm cap} =  16\pi \, \frac{ \alpha^2 k_\chi^2}{ m_\chi^2  w(0)^4} L_{\chi\phi}^{\rm cap}(m_\phi) \simeq \\
& \left(\frac{10\mbox{ GeV}}{m_\chi}\right)^2 \left(\frac{k_\chi}{\bar k_\chi^\Omega}\right)^2\left(\frac{L^{\rm cap}_{\chi\phi}}{\bar L_{\chi\phi}^{\rm cap}}\right) \cdot 9.6\times 10^{-25}\mbox{ cm}^2\ ,
\end{split}
\end{equation}
where $w(0)\simeq 1400$ km/s is the infalling velocity of the DM particles at $r=0$. %and $k_{10}^\Omega = \bar k_\chi^\Omega \sqrt{10}\simeq 0.24$ is the lower value of the parameter $k_\chi$ such that the constraint $k_\chi \geq k_\chi^\Omega$ for $m_\chi=10$ GeV holds. 
Here $L_{\chi\phi}^{\rm cap}=\ln(m_\chi w(0)/m_\phi)$ is the Coulomb logarithm for the self-capture process and $\bar L_{\chi\phi}^{\rm cap}\simeq 8.4$ is its value  for $m_\phi=10$ keV. Since the self-capture cross section, in the long-range regime  is very big, we have also to  consider the capture of DM particles in the halo by other DM particles that have already been captured within the Sun. This effect  can lead to exponential growth in the number of captured dark matter particles as a function of time until the number of particles captured has become so large that the star is optically thick to DM. As pointed out by~\cite{2010PhRvD..82h3509T}, the number of DM particles captured, including the self-interactions, is then given by
\begin{equation}\label{Nself}
\left\{\begin{array}{l}
\displaystyle N^{\rm w}_\chi(t)=\frac{\Gamma_{\rm cap}}{\Gamma_\chi}(e^{\Gamma_\chi t}-1) \quad \mbox{for } t\leq \hat t, \\
\displaystyle N^{\rm w}_\chi(t)=\left(\Gamma_{\rm cap} + \hat\Gamma_\chi \right)\left(t-\hat t\right)+N^{\rm w}_\chi(\hat t) \quad \mbox{for } t> \hat t,
\end{array}\right.
\end{equation}
where $\hat t$ is the  time at which the Sun becomes optically thick to DM,  $N^{\rm w}_\chi(\hat t) = \pi r_\chi^2(m_\chi)/\sigma_\chi^{\rm cap}$ is the critical number of DM particles captured due to the self-interaction and  $\hat \Gamma_\chi = \Gamma_\chi N^{\rm w}_\chi(\hat t) \simeq (10 \mbox{ GeV}/m_\chi)^2 \cdot 5.5 \times 10^{26}$ s$^{-1}$ is the critical rate. In Eq.~\eqref{Nself} the capture and self-capture rates, in the long-range regime, can be  estimated by the following analytic equations:
\begin{equation}\label{GammaT}
\begin{split}
&\Gamma_{\rm cap}=\sqrt{\frac32}\frac{\rho_\chi}{m_\chi}  \lim_{r_\phi\gg1} \sigma_T^{\rm cap}  \frac{v_{\rm esc}^2(R_\odot)}{v_0}N_\odot \langle \phi \rangle \frac{{\rm erf}{(\eta)}}{\eta} \simeq  \\
& \left(\frac{10\mbox{ GeV}}{m_\chi}\right) \left(\frac{\gamma_\phi}{10^{-9}}\right)^2\left(\frac{L^{\rm cap}_\phi}{\bar L_\phi^{\rm cap}}\right) \cdot 2.5 \times 10^{26} \, \rm s^{-1} \ ,
\end{split}
\end{equation}
and
\begin{equation}\label{GammaDM}
\begin{split}
&\Gamma_\chi=\sqrt{\frac32}\frac{\rho_\chi}{m_\chi}  \lim_{r_\phi\gg1} \sigma_\chi^{\rm cap}\frac{v_{\rm esc}^2(R_\odot)}{v_0} \langle \phi_\chi \rangle \frac{{\rm erf}{(\eta)}}{\eta} \simeq \\
& \left(\frac{10\mbox{ GeV}}{m_\chi}\right)^3 \left(\frac{k_\chi}{\bar k_\chi^\Omega}\right)^2\left(\frac{L^{\rm cap}_{\chi\phi}}{\bar L_{\chi\phi}^{\rm cap}}\right) \cdot 2.9 \times 10^{-17} \, \rm s^{-1} \  .
\end{split}
\end{equation}
with $\langle \phi \rangle \simeq 3.29$,  $\langle \phi_\chi \rangle \simeq 5.13$ and $\eta\simeq 1.29$ (see e.g.~\cite{Zentner:2009is}). The critical time $\hat t$ can then be obtained from the first line of Eq.~\eqref{Nself} by substituting $t$ with $\hat t$ and it reads explicitly 
\begin{equation}\label{tc}
\begin{split}
&\hat t = \frac1{\Gamma_\chi}\, \ln \left(1+\Delta_N  \right) \simeq \\
& 0.24 \, t_\odot \cdot \left(\frac{m_\chi}{10\mbox{ GeV}}\right)^3 \left(\frac{\bar k_\chi^\Omega}{k_\chi}\right)^2\left(\frac{\bar L_{\chi\phi}^{\rm cap}}{L^{\rm cap}_{\chi\phi}}\right) \ln \left(1+\Delta_N  \right) , 
\end{split}
\end{equation}
where $\Delta_N = \hat \Gamma_\chi / \Gamma_{\rm cap}$ and $t_\odot \simeq 4.567 \times 10^9$ years. 

Imposing now the constraint $k_\chi \gtrsim \bar k_\chi^\Omega \sqrt{m_\chi/{\rm GeV}}$ in order that the symmetric component of $\chi$ is depleted in the early universe, one can check from the equations above that $\hat t \ll  t_\odot $  for most of the parameter space considered in our analysis. In this case we can easily estimate which is the present time ratio  between  the number of DM particles captured including the self-interaction (the second line of Eq.~\ref{Nself}) and those captured neglecting this important effect (the analytic standard solution without the self interaction is $N_\chi(t_\odot)=\Gamma_{\rm cap} t_\odot$). It reads
\begin{equation}\label{Nselfratio}
 \frac{N^{\rm w}_\chi(t_\odot)}{N_\chi(t_\odot)}\simeq 1+\Delta_N \ ,
\end{equation}
where the function $\Delta_N$ has been defined in Eq.~\eqref{tc} and writes explicitly 
\begin{equation}\label{DeltaN}
\begin{split}
& \Delta_N= \frac{ r_\chi^2(m_\chi) \, m_p^2 \bar w^4}{4 N_\odot \alpha^2 \gamma_\phi^2 L_\phi^{\rm cap}(m_\phi)}\frac{ \langle \phi_\chi \rangle}{ \langle \phi \rangle} \simeq \\
&  2.2 \cdot  \left(\frac{10\mbox{ GeV}}{m_\chi}\right) \left(\frac{10^{-9}}{\gamma_\phi}\right)^2\left(\frac{\bar L^{\rm cap}_{\phi}}{L_{\phi}^{\rm cap}}\right)  \ .
%\end{split}
\end{split}
\end{equation}
As is apparent,  since the luminosity carried by the DM particles is proportional to the number density of them,  the increase in the number of DM particles captured due to the self-interaction described in Eq.~\eqref{Nselfratio} would go in the direction of increasing the effects pointed out in this paper. 

However, there is a countervailing effect associated with quite large DM self interaction. Indeed, when the Sun becomes optically thick to its own DM the mean free path  become in general shorter. Therefore the luminosity carried by DM particles in the conductive regime ($K_\chi \ll 1$) is reduced and in turn  the effects pointed out in this paper will then decrease. In order now to quantify this effect, the first thing that one has to define is of course the average mean free path of the DM particles in the Sun. Assuming now a Maxwell-Boltzmann distribution for the DM particles such that $n^{\rm w}_\chi(r)=  n_0^{\rm w}\,\mbox{exp}[-r^2/r_\chi^2]$ where $n_0^{\rm w} = N_\chi^{\rm w}/(\pi^{3/2}r_\chi^3)$, it writes
\begin{equation}\label{eq-meanfreepath-SI}
\begin{split}
& l^{\rm w}_\chi = l_\chi \left(1+l_\chi \, \langle n_\chi \rangle  \cdot \lim_{r_\phi\gg1} \sigma_\chi^{\rm cap} \right)^{-1}  \ ,
\end{split}
\end{equation}
where 
\begin{equation}
\langle n_\chi \rangle=\frac1{R_\odot} \int_0^{R_\odot}\hspace{-.15cm} {\rm d} r  \, n_\chi(r) = \frac{N_\chi^{\rm w}}{2\pi} \frac{{\rm erf}\left(R_\odot / r_\chi \right)}{R_\odot \, r_\chi^2 }\ ,
\end{equation}
is the average number density of DM particles in the Sun. In analogy with Eq.~\eqref{Nselfratio}, we can then estimate  which is the present time ratio  between  the average mean free path including the self-interaction (Eq.~\eqref{eq-meanfreepath-SI}) and the  one neglecting this important effect (the mean free path without the self-interaction $l_\chi$ is given in Eq.~\eqref{eq-meanfreepath-LR}). By means now of the second line of Eq.~\eqref{Nself}, the total number of DM captured due to the self-interaction can be written as $(1+\Delta_N)\Gamma_{\rm cap} t_\odot$ and therefore
\begin{equation}\label{lselfratio}
\begin{split}
\frac{l_\chi^{\rm w}}{l_\chi}= \left(1+\left(1+\Delta_N\right)\Delta_l \right)^{-1} \ ,
\end{split}
\end{equation}
where the function $\Delta_l$ is given by
\begin{equation}\label{Deltal}
\begin{split}
& \Delta_l=  \frac{\Gamma_{\rm cap}t_\odot}{2\pi} \frac{{\rm erf}\left(R_\odot / r_\chi \right)}{R_\odot \, r_\chi^2 \langle n_b \rangle } \lim_{r_\phi\gg1} \frac{\sigma_\chi^{\rm cap}}{\sigma_T^{\rm tra}} \simeq  \\
&  0.054 \cdot \left(\frac{10\mbox{ GeV}}{m_\chi}\right)^4 \left(\frac{k_\chi}{\bar k_\chi^\Omega}\right)^2\left(\frac{L_l }{\bar L_l} \right)  \ .
\end{split}
\end{equation}
Here $L_l = L_{\chi\phi}^{\rm cap} L_\phi^{\rm cap} / L_\phi^{\rm tra}$ and $\bar L_l \simeq 12$ is its value for $m_\chi = 10$ GeV and $m_\phi = 10$ keV. It is worth noticing that the transport is mainly conductive once the constraint $k_\chi \gtrsim \bar k_\chi^\Omega \sqrt{m_\chi/{\rm GeV}}$ is imposed.

Having now Eqs.~(\ref{Nselfratio},\ref{lselfratio}) at our disposal, we can finally  estimate the  effect of the self-interaction on the Sun's sound speed profiles. Indeed since the  Sun's sound speed profiles are related with the effective luminosity carried by DM particles, in the conductive regime the ratio $\mathcal Y_{\rm SI} \equiv  n_\chi^{\rm w}l_\chi^{\rm w} /  n_\chi l_\chi$ is a good estimator which can be used to quantify the effect of the self-interaction in our results presented in the next section. It can be written in a compact form as
\begin{equation}\label{Results-SI}
\begin{split}
\mathcal Y_{\rm SI} = \left(\frac1{1+\Delta_N} + \Delta_l \right)^{-1}  \ .
\end{split}
\end{equation}
Considering now a fiducial model with $m_\chi = 10$ GeV, $m_\phi = 10$ keV and $\gamma_\phi = 10^{-9}$ we have found that $1.17 \gtrsim \mathcal Y_{\rm SI} \gtrsim 1$ for $k_\chi^\Omega \gtrsim  k_\chi \gtrsim 1.13 \, k_\chi^\Omega $. The reason of that relies on the fact that there is a compensation in the Sun's sound speeds profiles between the larger number of the DM captured due to the self-interaction and the shorter mean free path. The range of the parameter $k_\chi$ in which $\mathcal Y_{\rm SI}\sim1$ does not depends too much on $\gamma_\phi$ and $m_\phi$, while it scales as $m_\chi^{-3}$ with the DM mass.  

On the other hand, if $k_\chi \gtrsim 1.5 \, k_\Omega$, the ratio $ \mathcal Y_{\rm SI} \simeq 1/\Delta_l$. In that case there is not an efficient compensation and therefore $\mathcal Y_{\rm SI}$ can reach even small value\footnote{For example, considering the maximal value of $k_\chi = 1/\sqrt{\alpha}\simeq 11.7$ allowed by  perturbation theory,  we get $\mathcal Y_{\rm SI} \simeq 7.5 \times 10^{-4}$ for the benchmark model.}.  In view of these facts we can conclude that is a good approximation neglecting the self-capture rate in Eq.~\eqref{eq-N}  if $k_\chi \sim k_\chi^\Omega$. Thanks to this constraint we can  fix $k_\chi = k_\chi^\Omega$ and directly present the final results  in terms of the kinetic mixing parameter $\epsilon_\phi = \gamma_\phi / k_\chi^\Omega$. %The error of this choice can be estimated from the relation abs[$\left(\mathcal Y_{\rm SI}-1\right)]^{1/2}$.

}

\begin{figure*}
\centering
\includegraphics[scale=.8]{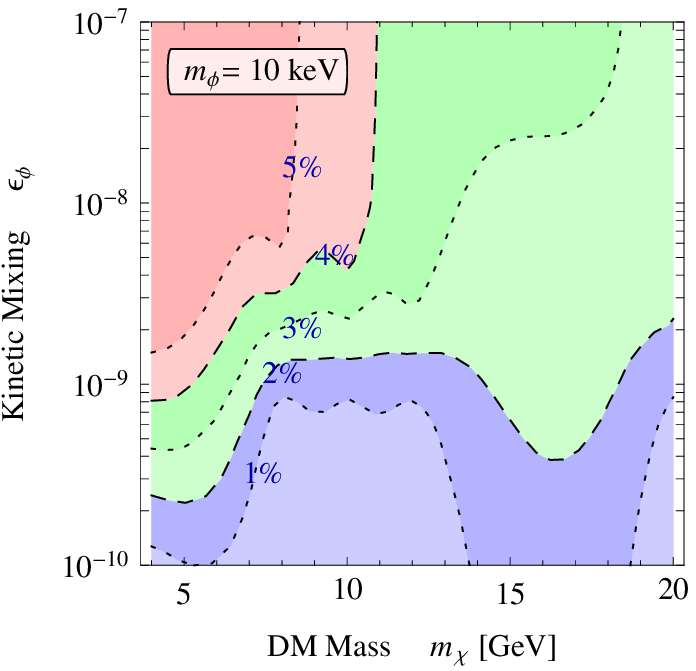} \
\includegraphics[scale=.8]{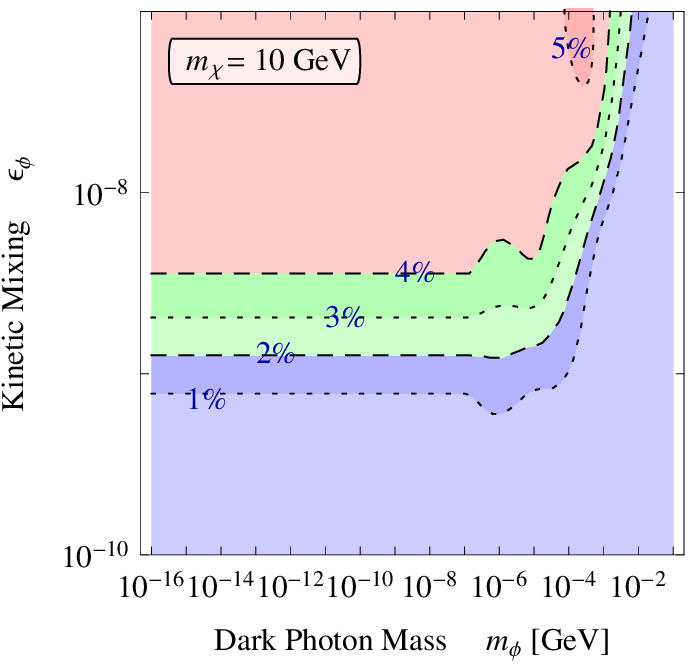} \
\includegraphics[scale=.8]{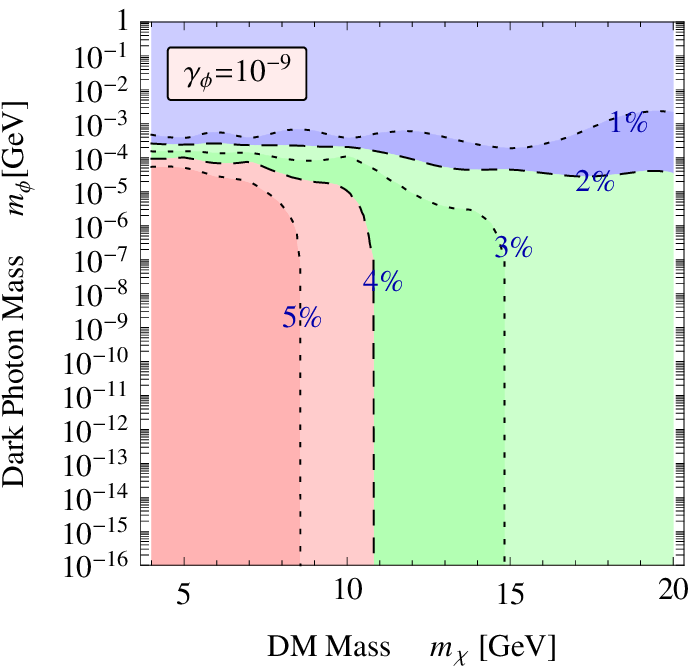}
\caption{
The maximum sound speed difference $\Delta c^2_{\rm max}=\max{\left[(c_{\rm mod}^2-c_{\rm ssm}^2)/c_{\rm ssm}^2\right]}$ in the full parameter space  of DMLRI  models ($\epsilon_\phi,m_\chi,m_\phi$) once the constraint $k_\chi=k_\chi^\Omega(m_\chi)$ is imposed. {\it Left panel:} Parameter space projected in the ($m_\chi, \epsilon_\phi$) plane keeping fix $m_\phi=10$ keV; {\it Central panel:} Parameter space in the ($m_\phi, \epsilon_\phi$) plane considering $m_\chi=10$ GeV; {\it Right panel:} Projection of the parameter space in the ($m_\chi,  m_\phi$) plane for a fix $\gamma_\phi=\epsilon_\phi k_\chi^\Omega(m_\chi) = 10^{-9}$. In all panels the red(blue) areas individuate the regions of the parameter space where $\Delta c_{\rm max}^2>$ 4\%($\Delta c^2_{\rm max}<2\%$) while those in light green  refer to the regions where the agreement with helioseismic data is better than the SSM (2\% $<\Delta c_{\rm max}^2<$ 4\%). All the DMLRI models in the red regions are excluded since they produce a large impact on the Sun's core sound speed profile. The DM halo in the Galaxy has been assumed in the form of an isothermal sphere with local energy density $\rho_\odot=0.38$ GeV/cm$^3$ and velocity dispersion $v_0=220$ km/s.
}
\label{fig:3}
\end{figure*}

%%%%%%%%%%%%%%%%%%%%%%%%%%%%%%%%%%%%%%%%%%%%%%%%%%%%%%%%%%%%%%%%%%%%%%%%%%%
\section{Discussion}
\label{Sec:4}
%%%%%%%%%%%%%%%%%%%%%%%%%%%%%%%%%%%%%%%%%%%%%%%%%%%%%%%%%%%%%%%%%%%%%%%%%%%

The impact of DM in the Sun is studied by inferring the modifications 
that DM causes to the Sun's structure and to  the solar observables. 
In the following, the standard solar model~\citep[SSM; e.g.,][]{1993ApJ...408..347T, 2013arXiv1308.3346L} 
is used as our model of reference, which predicts solar neutrino fluxes 
and helioseismology data consistent with current measurements. 
The excellent agreement obtained between theory and observation 
results from the combined effort between the fields of helioseismolgy 
and  solar modelling, a collaboration extended by several decades
which lead to a high precision description of physical processes
present inside the Sun~\citep{2011RPPh...74h6901T, 2012RAA....12.1107T}.  
This was very relevant in the case of the physical processes related with microscopic physics, including 
the equation of state, opacities, nuclear reactions rates, and microscopic diffusion of helium and heavy elements. 
A detailed discussion about current predictions of the SSM and their uncertainties can be found in the
literature~\citep[e.g.,][]{1993ApJ...408..347T, 2009ApJ...705L.123S, 2010ApJ...713.1108G, 2010ApJ...715.1539T, 2013ApJ...765...14L, 2013arXiv1308.3346L, Lopes:2013uxa}.   

\medskip

{In Fig.~\ref{fig:2} we compare the sound speed profile of SSM with
the sound speed computed 
by an inversion technique from helioseismology data~\citep{1997SoPh..175..247T, 2009ApJ...699.1403B}.  
The green square dots correspond 
to the relative sound speed difference  $\Delta c_{obs}=(c_{\rm obs}^2- c_{\rm ssm}^2)/c_{\rm ssm}^2$,
where $c_{\rm ssm}$ and $c_{\rm obs}$ are the sound speed from SSM and helioseismic data.  
$\Delta c_{obs}$   is smaller than 2\% throughout the solar interior, 
above 20\% and below 90\% of the Sun's radius.
Although agreement between $c_{\rm ssm}$ and $c_{\rm obs}$ is very good,
a discrepancy remains between the present SSM and helioseismic data, from which there is no 
 obvious solution~\citet{2011RPPh...74h6901T}. It is worth noticing    
that the quality of the sound speed inversion is highly reliable, as most of the helioseismic data 
has a relative precision of measurements larger than $10^{-4}$. Contrarily, 
in the Sun's inner core below $0.2\;{\rm R_\odot}$, the seismology data available is quite sparse
and consequently the sound speed inversion is less reliable (cf. Fig.~\ref{fig:2}).  
As pointed out by~\citet{2011RPPh...74h6901T} the inversion of the sound speed profile 
in the Sun's inner  core is limited by the low number acoustic frequencies measured
(see Table 1 in~\citet{2012RAA....12.1107T} and references therein), 
as well as by the weak sensitivity of the eigenfunctions of global acoustic modes 
to the structure of the Sun's core. This difficulty can only 
 be overcome with the positive detection of gravity modes.    
Equally, in the most external layers of the Sun, the inversion of the sound
speed profile is not possible, mainly due to the fact that the inversion 
technique breaks down (acoustic oscillations are no-longer adiabatic), 
as a complex interaction occurs between convection, magnetic fields 
and acoustic oscillations~\citep{Gough:2012ebb,2001MNRAS.322..473L}.  
   
Accordingly, for the purpose of this study, we choose to consider the theory-observation 
uncertainty to be of the order of 4\% rather than 2\%. In the remainder of the article
we will refer to this value as the {\it SSM uncertainty}, meaning the undistinguished uncertainty 
related to the physical processes of the SSM or helioseismogy sound speed inversion.}
 
\medskip 
The DMLRI solar models were obtained in an identical manner to the SSM,
by adjusting the initial helium  $Y_{i}$ and the mixing length parameter $\alpha_{MLT}$ 
in such a way that at the present age ($4.6 \;\; {\rm Gyear}$), 
these solar models reproduced the observed values of the  mass, radius and luminosity
of the Sun, as well as the observed photospheric abundance ratio $(Z/X)_{\odot}$, where $X$ and $Z$   
are the mass fraction of hydrogen and the mass fraction of elements heavier than helium, respectively.
{Fig.~\ref{fig:2} shows a comparison between SSM and different DMLRI  models. 
The different continuous lines correspond to the squared sound speed difference    
$\Delta c_{\rm mod}^2=(c_{\rm mod}^2- c_{\rm ssm}^2)/c_{\rm ssm}^2$ where $c_{\rm mod}$
is the sound speed of DMLRI solar models. 
These  models are obtained for a fiducial value of $\gamma_\phi=10^{-9}$
and different values of $m_\chi$ and $m_\phi$. The most important point about Fig.~\ref{fig:2}
 is the fact that there are some DMLRI solar models that can resolve the current discrepancy
with helioseismology data, as $\Delta c_{\rm mod}^2$ reproduces 
the observed discrepancy   $\Delta c_{\rm obs}^2$. }

{In DMLRI models,  the DM impact is most visible in the core of the
star where the DM particles accumulate. However,
because the solar models are required to have the current observed 
values of radius and luminosity,
a decrease of the production of nuclear energy in the Sun's core
due to the reduction of the central temperature 
(caused by the thermalisation of DM with baryons),  
is compensated by an increase of the sound speed in the radiative region.
In Fig.~\ref{fig:2} we show an illustrative DMGRI solar model
with benchmark parameters: $m_\chi=10\;{\rm GeV}$, $m_\phi=10\;{\rm keV}$  and $\gamma_\phi=10^{-9}$
({\it black curve}).
Moreover, all the DMLRI solar models have an identical 
impact behaviour on the solar structure, however, 
based upon the parameters $m_\chi$ and $m_\phi$ it is possible to distinguish
three sets of models: 
$i)$ DMLRI models for which the squared sound speed difference is  larger than the {\it SSM uncertainty} ({\it red curves});
$ii)$ DMLRI models for which the agreement with the helioseismic data 
is better than the current SSM ({\it green curves}); $iii)$  DMLRI models for which the squared sound speed difference is smaller than the {\it SSM uncertainty} ({\it blue curves}). 
Although, there is a large set of  DMLRI models ({\it red } and {\it green curves}) for which
$\Delta c_{\rm mod}^2$ tends to agree $\Delta c_{\rm obs}^2$, as for those models for which 
the central value of $c_{\rm mod}$ varies more than $4\%$,  the central temperature will 
change for an identical amount. Consequently, the solar neutrino fluxes
of these models become strongly in disagreement with the current neutrino flux observations~\citep[e.g.,][]{2012ApJ...752..129L,2012ApJ...757..130L}.
Therefore, we make the conservative option in this preliminary study 
to only consider models for which the central temperature does not 
change very much from the  SSM ({\it green curves} DMLRI models). }

Fig.~\ref{fig:3} shows the maximum sound speed difference $\Delta c^2_{\rm max}=\max{\left[(c_{\rm mod}^2-c_{\rm ssm}^2)/c_{\rm ssm}^2\right]}$ in the full parameter space  of DMLRI  models ($\epsilon_\phi,m_\chi,m_\phi$) once  the constraint $k_\chi=k_\chi^\Omega(m_\chi)$ is imposed. The represented percentages are performed for $r\le 0.3 \,{\rm R_{\odot}}$. On a more specific level in the right panel of Fig.~\ref{fig:3} we project the parameter space in the ($m_\chi,\epsilon_\phi$) plane by choosing $m_\phi=10$ keV. In the central plane the ($m_\phi,\epsilon_\phi$) plane is shown for $m_\chi=10$ GeV, while on the left  panel, the parameter space is projected in the ($m_\chi, m_\phi$) plane keeping fix $\gamma_\phi=\epsilon_\phi k_\chi^\Omega(m_\chi) =10^{-9}$. Once $\Delta c^2_{\rm max}$ is larger than the {\it SSM uncertainty}  it is reasonable to exclude all DMLRI models, since they produce a large impact on the Sun's core sound speed profile ({\it red regions}). On the other hand, if $\Delta c^2_{\rm max}$ is in the range $2\%-4\%$ the agreement with the helioseismic data, as  commented upon above, is improved ({\it  green regions}). For instance, we find that DM particles with a mass in the range 4 GeV--8.5 GeV coupled with ordinary baryons via a kinetic mixing parameter $\epsilon_\phi$ bigger than $5\times 10^{-9}$, produce a very large impact on the Sun's core in the long--range regime ($m_\phi$ smaller than a few MeV). Therefore they can be excluded as possible DM candidates. On the other hand, we can see that DM particles with a mass of the order of $10\;{\rm GeV}$,  a kinetic mixing parameter $\sim 10^{-9}$ and a mediator with a mass smaller than a few MeV  improve  the agreement between the best solar model and the helioseismic data. This is quite interesting since direct DM searches experiments, as we shall see in the next section, either do not, or barely exclude,  these kinds of DM models with long--range interactions with baryons. {Furthermore, it is worth noticing that for very light dark photon ($m_\phi$ smaller than few keV), and for $\epsilon_\phi \gtrsim 3.33\times 10^{-9}/k_\chi^\Omega(m_\chi)$ the transport occurs in the conductive regime. In this case, as we can see in the figure, our results are independent on $m_\phi$, because the effective luminosity carried by DM particles  in the conductive regime is proportional to $n_\chi l_\chi \propto \sigma_T^{\rm cap}/\sigma_T^{\rm tra}$. This is important because, as we will see in the next section, the complementary constraints coming from supernova observations and astrophysical or cosmological arguments are very week for very light mediators. 

}

\begin{figure*}[t]
\centering
\includegraphics[scale=0.815]{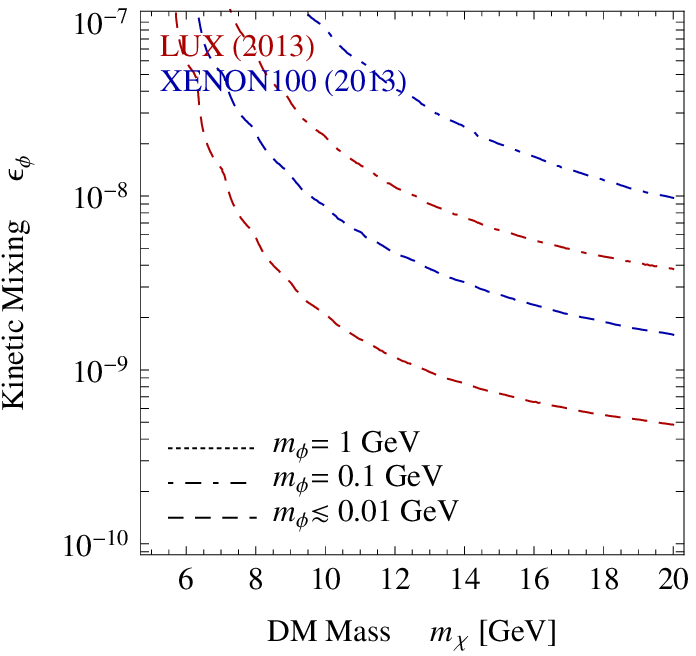} \quad
\includegraphics[scale=0.815]{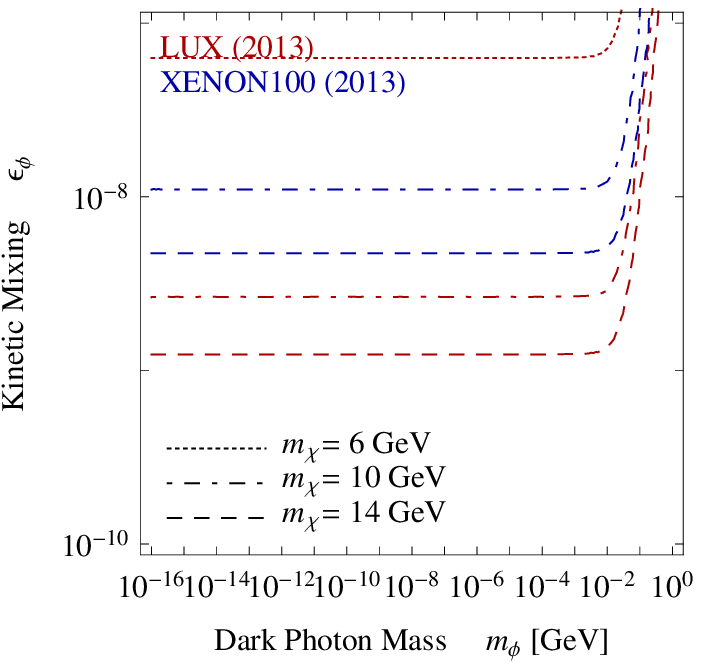} \quad
\includegraphics[scale=0.815]{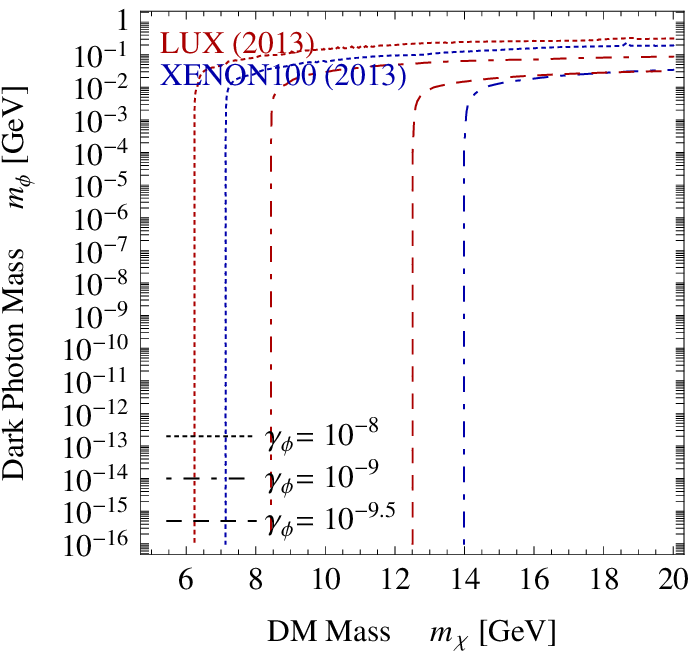}
\caption{
Direct detection constraints in the relevant parameter space of DMLRI models once the constraint $k_\chi=k_\chi^\Omega(m_\chi)$ is imposed. {\it Left panel:} Parameter space projected in the ($m_\chi, \epsilon_\phi$) plane; {\it Central panel:} Parameter space in the ($m_\phi, \epsilon_\phi$) plane; {\it Right panel:} Projection of the parameter space in the ($m_\chi,  m_\phi$) plane. In all panels, the dark blue(red) lines refer to the constraints coming from \XENON(\LUX), while the different hatching (dotted, dashed-dotted, dashed) indicates the three values of the third direction in the parameter space we kept fixed ($m_\phi=$(1, 0.1, $\leq$ 0.01) GeV on the left panel, $m_\chi=$(6, 10, 14) GeV on the central panel and $\gamma_\phi=\epsilon_\phi k_\chi^\Omega(m_\chi)=(10^{-8}, 10^{-9}, 10^{-9.5})$ on the right panel). Like in Fig.~\ref{fig:3}, the bounds are computed by assuming an isothermal halo with $\rho_\odot=0.38$ GeV/cm$^3$ and $v_0=220$ km/s.} 
\label{fig:4}
\end{figure*}

\section{Complementary Constraints}\label{CConstraints}
%%%%%%%%%%%%%%%%%%%%%%%%%%%%%%%%%%%%%%%%%%%%%%%%%%%%%%%%%%%%%%%%%%%%%%%%%%%
In this section, we present the complementary  constraints which are relevant for DMLRI models. {A first class of them come from terrestrial direct detection experiments. In particular, since the interactions considered in our work is spin-independent, we only compute the constraints coming from the \XENON\  and \LUX\ experiments. The statistical analysis used to treat the datasets can be found in \citep{DelNobile:2013sia} where the authors have provided a complete sets of numerical tools for deriving the bounds in direct searches in a completely model independent way. Without entering  in the details of this work we can compute the bounds in the relevant parameter space of our model following the main steps summarized in Sec.~6 of  \citep{DelNobile:2013sia}. In particular, as pointed out in  the first three steps (1a-1c), we have to identify the non-relativistic operator and its coefficient associated with a Yukawa-type interaction given in Eq.~\eqref{Vr}. Since this interaction is spin independent the operator is simply the identity, while the non-relativistic coefficient is given by $c_{\rm Y}^p = 16 \pi \alpha \gamma_\phi m_p m_\chi /(q^2+m_\phi^2)$ (see e.g.~\citep{Panci:2014gga} for details). Thanks now to ready-made scaling functions provided in the webpage of \citep{DelNobile:2013sia}, the bounds on the parameter space of our model can be obtained by following the last two steps (2a-2b).} In Fig.~\ref{fig:4} we show the constraints in the relevant parameter spaces of our model. As we did in the previous section, we project the parameter space in the ($m_\chi, \epsilon_\phi$) plane (left panel), ($m_\phi, \epsilon_\phi$) plane (central panel) and ($m_\chi, m_\phi$) plane (right panel). The dark blue (red) lines refer to the constraints coming from \XENON(\LUX), while the different hatching (dotted, dashed-dotted, dashed) indicates the three values of the third direction in the parameter space  that we kept fixed ($m_\phi=$(1, 0.1, $\leq$ 0.01) GeV on the left panel, $m_\chi=$(6, 10, 14) GeV on the central panel and $\gamma_\phi=\epsilon_\phi k_\chi^\Omega(m_\chi)=(10^{-8}, 10^{-9}, 10^{-9.5})$ on the right panel). The areas of the parameter space above the first and second plots, and the ones below the third plot are excluded. Similar constraints can also be found in Ref.~\citep{Kaplinghat:2013yxa}. We can see that in the long--range regime (mediator masses below roughly 10 MeV), the constraints becomes independent of $m_\phi$. Indeed in this case, the interaction is Rutherford-like and therefore the differential cross section in Eq.~\eqref{diffCC}, which sets the normalization of the total number of events in  certain experiments, solely depends  on the exchanged momentum $q$. By virtue of this fact, we can just use the left plot of Fig.~\ref{fig:4} and compare it directly with the results shown  in Fig.~\ref{fig:3}. We can see that in the relevant DM mass range that affects helioseismic data (4--20) GeV, the constraints, coming from the \LUX\ experiments, exclude DMLRI models that are coupled with baryons through kinetic mixing parameters larger than roughly ($\gtrsim 10^{-7}, 5\times 10^{-10}$). Therefore, as is apparent,  long-range spin-independent DM-baryon interactions can easily improve  the agreement between the best solar model and the helioseismic data without being excluded by direct detection experiments for DM masses in the range (4-12) GeV.  Indeed, as we have already pointed out, the Sun is an ideal experiments for DM models which posses an enhanced cross section with baryons for small  momentum exchanges.

\medskip
{A second class of complementary constraints that only depends on the properties of the dark photon (kinetic mixing parameter $\epsilon_\phi$ and its mass $m_\phi$) are those coming from supernov\ae\ observations and beam dump neutrino experiments. These are presented, for instance, in the $(m_\phi,\epsilon_\phi)$ plane in Fig.~6  ($m_\phi > 10^{-3}$ GeV) and Fig.~7  ($m_\phi <10^{-3}$ GeV) of \citep{Essig:2013lka}. We can see that the most stringent constraints are the ones coming from supernova observations (namely the energy loss observed from SN1987a) and from astrophysical, or cosmological arguments: they indeed constrain  relatively small kinetic mixing parameters ($\epsilon_\phi$ above roughly ($10^{-11}-10^{-10}$)) and dark photon masses in the range ($10^{-9} \lesssim m_\phi \lesssim 0.2$) GeV. Therefore they ruled out all the parameter space above $m_\phi\gtrsim 10^{-9}$ GeV shown in the central panel of Fig.~\ref{fig:3}. On the other hand for very light messengers, they become extremely weak ($\epsilon_\phi \gtrsim 10^{-8}$) being the direct production of dark photons forbidden by kinematical reasons.  In that case  all the parameter space will be then reopened since, as we  commented upon in the last paragraph of Sec.~\ref{Sec:4}, the amplitude of the Sun's sounds speed profiles is independent on $m_\phi$ (see e.g. the central panel of Fig.~\ref{fig:3}). }

\medskip
A third class of constraints which instead solely depend on the characteristics on the dark sector itself are those coming from  self--interactions. Indeed, since for this type of model, the DM-DM scattering is not suppressed by $\epsilon_\phi$,  the self--interactions, especially in the long--range limit, can easily reach large values, affecting the dynamics of virialized astrophysical objects. In Refs.~\citep{Tulin:2013teo,Kaplinghat:2013yxa},  the constraints on $m_\phi$ coming from the observations of a few elliptical DM halos have been presented. These constraints are in general very stringent for DM masses below 10 GeV: in this case, in fact, the time over which  energy is transferred in the system is extremely rapid and therefore a spherical DM halo tends to form in contradiction with observations. In particular, dark photon masses below roughly 100 MeV are excluded. However, the derivation of this last class of constraints is very uncertain, both from the theoretical and experimental side, because we
do not really know how virialized astrophysical objects are formed in the presence of a DM sector with long--range interactions.  Indeed, since the self-interaction needed to change their dynamics is in general  of the order of the Thompson scattering ($\sigma_{\rm em}\simeq 10^{-24}$ cm$^2$), from trivial  dimensional analysis of such large cross sections, the following rough 
estimate yields that   the self-interaction is of  a long--range type in most of the virialized astrophysical objects under the assumption that the DM-dark photon coupling is of the order of $\alpha$.  % For bigger value of $\alpha$ the interaction can be of a contact-type but one has to assume that the dark sector is strongly coupled with the dark photon. 
In this case,  the probability to radiate a dark photon in the scattering process is different from zero and therefore 
  it might  well be  possible that  the DM sector  is dissipative  just like our sector. The time over which energy is transferred in the system is no longer a good indicator, since  the relevant quantity that describes the dynamical evolution of the system is now the cooling time: in particular, for a DM sector  composed of heavy and light species, the dissipation time due to the soft emission of dark photons (dark bremsstrahlung)  can be smaller than the age of the virialized astrophysical objects \citep[see e.g.][]{Fan:2013tia, Fan:2013yva,  McCullough:2013jma}. In this case, the system is no longer stable and in general it starts to collapse. By virtue of this fact, we do not consider this last class of constraints, since a more dedicated and careful analysis  also involving  numerical simulations is clearly  needed.

%%%%%%%%%%%%%%%%%%%%%%%%%%%%%%%%%%%%%%%%%%%%%%%%%%%%%%%%%%%%%%%%%%%%%%%%%%%
\section{Conclusion and Summary}
%%%%%%%%%%%%%%%%%%%%%%%%%%%%%%%%%%%%%%%%%%%%%%%%%%%%%%%%%%%%%%%%%%%%%%%%%%%

We have examined how DM-baryon long--range interactions can affect the Sun's sound speed radial profile. 
The phenomenological approach used in our analysis lets  us explore all the parameter space from 
the contact regime to the long-range regime.  We find that DM particles lighter than 8.5 GeV coupled with 
ordinary baryons through a kinetic mixing parameter $\gamma_\phi$ bigger than $5\times 10^{-9}$, 
produce a very large impact on the Sun's core in the long--range regime 
($m_\phi$ smaller than few MeV). Therefore they can be excluded as possible DM candidates.  
However, solar models for which the DM particle has a mass of $10\;{\rm GeV}$
and the mediator a mass smaller than $1\;{\rm MeV}$ improve 
the agreement  with helioseismic data. 
Nevertheless, when the mass of the dark photon is larger
than $10\;{\rm MeV}$ the impact on the Sun's structure is very small, being 
in this case the interaction of a contact-type (standard spin-independent picture).
In particular, the results obtained here reveal that DM models featuring a long-range 
interaction with ordinary matter can affect the sound-speed radial profile and in turn probably
solve the so-called solar abundance problem without being excluded by terrestrial experiments 
(e.g. \LUX\ and \XENON) in which the scattering cross section is suppressed by the strong inverse dependence on $q$.
The Sun is in fact an ideal  DM detector for the type of particle considered here, 
since it measures the entire  nuclear recoil energy spectra in  the scatterings.
   
\medskip
To summarize, in this work we have obtained two main results. 
Firstly, for the first time, a DM-baryon velocity dependent 
total cross-section has been implemented  in  solar simulation software.
Secondly, but even more importantly, our  analysis shows that DM particles with a mass of 10 GeV and
a long--range  interaction with ordinary matter mediated by a very light mediator (below roughly  a few MeV), can have an impact on the Sun's sound speed profile without  violating the constraints coming from direct DM searches. {Our results are valid if the parameter $k_\chi$ which controls the self-interaction is of the order of $k_\chi^\Omega$. Larger values of this parameter, as we have shown in the last paragraph of Sec.~\ref{Sec:4}, can dramatically reduced the effective luminosity carried by DM particles in the Sun and therefore our results can not be applied in that case.}   Furthermore, as commented upon in Sec.~\ref{CConstraints}, it might well be possible that a dark sector with long--range forces can  dissipate a relevant amount of energy through the emission of  dark photons.  A dissipative dark sector is extremely interesting because it can offer a rich array of new ideas ranging from the detection of primordial dark radiation to new possibilities for  DM dynamics in virialized astrophysical objects. 
   
% % % % % % % % % % % % % % % % % % % % % % % % % % % % % % % % % % % % % % % % % %
\begin{acknowledgments}
We would like to thank Marco Taoso for useful discussions.  This work was supported by grants from "Funda\c c\~ao para a Ci\^encia e Tecnologia" 
and "Funda\c c\~ao Calouste Gulbenkian".  This research has been supported at IAP by  ERC project  267117 (DARK) 
hosted by Universit\'e Pierre et Marie Curie - Paris 6,   and at JHU by NSF grant OIA-1124403.
\end{acknowledgments}

\bibliographystyle{yahapj}

% % % % % % % % % % % % % % % % % % % % % % % % % % % % % % % % % % % % % % % % % % % % % % % % %
% % % % % % % % % % % % % % % % % % % % % % % % % % % % % % % % % % % % % % % % % %

\end{document}